\documentstyle[aps,prl,multicol,epsf]{revtex}
\renewcommand{\narrowtext}{\begin{multicols}{2} \global\columnwidth20.5pc}
\renewcommand{\widetext}{\end{multicols} \global\columnwidth42.5pc}

\renewcommand{\v}[1]{{\bf #1}}

\newcommand{\ba}{\begin{eqnarray}}
\newcommand{\ea}{\end{eqnarray}}
\newcommand{\be}{\begin{equation}}
\newcommand{\ee}{\end{equation}}

\newcommand{\Eq}[1]{Eq.~(\ref{#1})}

\begin{document}
\draft
\title{Staggered Currents in the Vortex Core}
\author{Jung Hoon Han$^1$, Qiang-Hua Wang$^{1,2}$, and Dung-Hai Lee$^1$}
\address{${(1)}$Department of Physics,University of California
at Berkeley, Berkeley, CA 94720, USA \\}
\address{${(2)}$ Physics Department and National Laboratory of Solid State Microstructures,\\ Institute for Solid State Physics, Nanjing University, Nanjing 210093, China\\}
\maketitle

\begin{abstract}
We study the electronic structure of the vortex core in the cuprates using  the U(1) slave-boson mean-field wavefunctions and  their Gutzwiller projection. We conclude that there exists local orbital antiferromagnetic order in the core near optimum doping. We compare the results with that of BCS theory and analyze the spatial denpendence of the local tunneling density of states.
\end{abstract}

\narrowtext

The high-T$_c$ superconductors are doped Mott insulators. Recently the possibility that the complex behavior of
high-T$_c$ systems is due to several competing orders came into focus. The contenders are the antiferromagnetic
order, $d$-wave superconducting order, and stripe order\cite{stripe}. Most recently yet another item has been
added to the above list - the staggered flux (SF) order\cite{ivanov,lt,clmn}. In the SF ground state there is a
current circulating around each plaquette in the copper-oxide plane giving rise to orbital-magnetic moments. Like
in a spin antiferromagnet, the orbital moment staggers from plaquette to plaquette.

Among different proposals of the SF order, Chakravarty {\it et al.} suggest that such order actually exists in
static form in the underdoped region of the high-T$_c$ phase diagram\cite{clmn}. So far there is no direct
experimental evidence for such a state. In another proposal, Ivanov, Lee, and Wen argue that the SF order appears
in dynamic rather than static form\cite{ivanov}. They substantiate their argument by showing that the
Gutzwiller-projected $d$-wave superconducting state possesses an equal-time, power-law SF correlation. Such
correlation is also seen in exact diagonalization of the t-J model\cite{leung}.  Based on SU(2) slave-boson mean-field
theory\cite{su2}, Lee and Wen further argued that the fluctuation of the SF order might slow down or even become
frozen in the core of a superconducting vortex\cite{leewen}.  They claim that the SF vortex core is more likely to
be found in the underdoped systems, because the energy difference between the SF state and the $d$-wave
superconducting state diminishes with underdoping.

In the following we present evidence for the existence of SF order in the vortex core using the U(1) slave-boson
mean-field wavefunction\cite{hl} and its Gutzwiller projection. Unlike Lee and Wen's expectation, however, we find
the strongest evidence for the SF vortex core {\it near optimum doping}, and a weakening evidence as the system
becomes either more overdoped or underdoped.

The above discrepancy triggers the following cautionary remark. In an unbiased mean-field theory of the t-J  model
with the Coulomb interaction between the holes, the following orders compete with each other: antiferromagnetism,
spin dimerization, staggered flux, and $d$-wave pairing. If no artificial constraint, such as preserving the
translation symmetry or suppressing antiferromagnetism, is imposed the mean-field ground state exhibits
antiferromagnetic order for small doping ($x\!<\!2\%$), stripe order for intermediate doping
($2\%\!<\!x\!<\!14\%$), and uniform $d$-wave superconducting order for high doping
($14\%\!<\!x\!<\!25\%$)\cite{hwl}. If the antiferromagnetism is omitted, the ground state for low doping shows
spin-dimerization. Such dimerized insulating states also appear in the insulating region of the stripe phase in
the intermediate doping\cite{sachdev}. If one forbids antiferromagnetism, stripes and dimerization all together,
then, aside from the $d$-wave superconducting phase\cite{kotliar}, the SF phase\cite{affmst} is the closest
contender for the ground state. In that case one expects that by frustrating the $d$-wave superconducting order
the SF order will be revealed. This leads naturally to the prediction that the SF order appears in the vortex core
or in the bulk when an external magnetic field suppresses the superconducting order\cite{leewen}. For sufficiently  low doping
this conclusion is likely to be invalid due to the presence of other competing orders.

We have previously carried out a numerical study of the vortex core within the
U(1) slave-boson mean-field theory\cite{hl}.
The Hamiltonian we considered is the t-J + Coulomb model. In standard notation the Hamiltonian reads
\begin{eqnarray}
H&=&-t\sum_{\langle ij \rangle} (b_jb^{\dag}_i f_{j\alpha}^\dag f_{i\alpha} \!+h.c.) +J\sum_{\langle ij\rangle}(\v
S_i\cdot\v S_j \! -\frac{1}{4}\!n_{i}n_{j})  \nonumber \\
&+& {\frac{V_c}{2}}\sum_{i\ne j} {\frac{1}{{r_{ij}}}}(n_{i}-\bar{n})
(n_{j}-\bar{n}) -\mu\sum_i n_i.
\label{h}
\end{eqnarray}
Equation (\ref{h}) is parametrized by two dimensionless ratios $t/J$ and $V_c/J$, where $V_c$ is the strength of
the Coulomb interaction at the nearest-neighbor distance. In Ref. \cite{hl} we have performed a partially-biased
mean-field search where the spin-antiferromagnetic order is disallowed. The result shows that depending on doping
there are two types of vortex core. The first type is insulating, and is favored at low doping. The second type is
metallic, favored at high doping. In the insulating core, the frustration of the $d$-wave superconducting order
gives rise to an insulating state known as the ``box phase''\cite{boxphase}. (This state is known to be equivalent
to the spin-dimer state at zero doping.) There is no orbital current flowing in this type of vortex core. Although
it went unnoticed by the authors at the time,  it turns out that in the metallic core SF order does
exist\cite{hl}. The purpose of this paper is to present in detail the SF state in the metallic vortex core. We
believe that the omission of antiferromagnetism is likely to have important effects on the structure of the
insulating core\cite{note}, but much less so on  the metallic one.
\begin{figure}
\centering
\hskip 0.1cm
\epsfxsize=5cm \epsfysize=5.5cm
\epsfbox{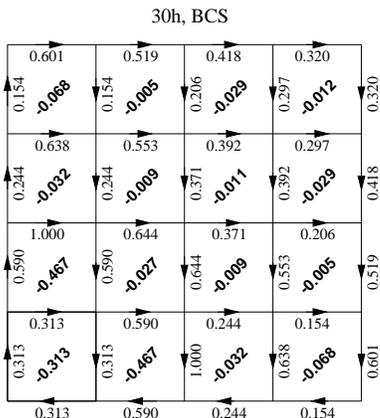}
\caption{The current pattern near the core of a BCS
vortex. The magnitude and direction of the bond current are shown. The numerals in the center of each plaquette is
the lattice curl of the bond current, $\phi (\v R)$, defined in \Eq{op}. The currents are normalized so that the maximum magnitude of the current is unity. The center of the vortex is shown in dark square, with one quadrant of the lattice in display.}
\end{figure}
Before showing the main results we first present the current pattern in the vortex core according to the $d$-wave
BCS theory\cite{note2}. In Fig. 1 the direction and magnitude (normalized by the maximum) of the bond current in
the core of a BCS $d$-wave vortex is presented. The result shown here is the self-consistent mean-field solution
for the t-J model upon ignoring the occupation constraint. We use $t/J=2$ and the average  number of electrons per
site equal to 0.88. It is clear that the current pattern is the diamagnetic pattern dictated by the vorticity. In
the figure the numerals inside each plaquette is the lattice curl of the bond current $J_{ij}$ defined respectively by
\begin{eqnarray}
& &J_{ij}=-i \langle c^{\dag}_j c_i \!-\!c^{\dag}_i c_j\rangle= -i \langle b^{\dag}_i b_j f^{\dag}_j f_i \rangle
\!+\! \mbox{h.c.} \nonumber \\ & &\phi(\v R)=(1/4)\!\sum_{\langle ij\rangle \in {\cal{P}}_R} \eta_{ij} J_{ij}.
\label{op} 
\ea 
Here $\v R$ labels the position of the plaquette ${\cal{P}}_R$ and $\epsilon_{ij}=\pm 1$ is chosen
so that the sum in \Eq{op} measures the circulation of $J_{ij}$ using right-handed rule. The fact that the sign of
$\phi (\v R)$ is the same for every plaquette attests to the absence of staggered current flow.

The results with the occupation constraint taken into account are shown in Fig. 2  for three doping concentration
$x=6\%, 12\%, 18\%$ with $t/J=2$ and $V_c/J=1.5$. The mean-field techniques are explained in Ref.\cite{hl}.
Here we simply stress that in this
calculation the slave-boson amplitude, fermion hopping and pairing amplitudes, and the Lagrange multipliers
are all determined self-consistently without any restriction. Once the self-consistent mean-field wavefunction is
obtained, we perform the Gutzwiller projection to remove double occupancy.
The mean-field results shown here is for $16\times 16$ lattice with an open boundary condition.
The central $10\times 10$ subset of the self-consistent mean-field parameters are used to
construct the input wavefunction for the Gutzwiller projection. The evaluation
of the current expectation value using the Gutzwiller projected wavefunction is achieved by
a variational Monte Carlo technique.
The results are obtained by running 5000 equillibration sweeps, and 10000 averaging
sweeps. Finally we symmetrize the bond currents according to the point group symmetry of the square lattice.

Figures 2(a)-(c) illustrate the mean-field results for $x=6\%, 12\%$, and $18\%$ respectively. Since the point
group symmetry is maintained in the solution, we only show one quadrant of the lattice with the center of the
vortex sitting in the middle of the dark square located at the lower left corner of each figure. The arrows in the
figure indicate the direction of the current flow. As in Fig. 1 we normalize the currents by their maximum value
(See Table I for a list of maximal currents). It turns out that for all cases the maximal current occurs one
plaquette away from the center. By inspection we conclude that the current pattern is the superposition of two
components: the diamagnetic flow dictated by the vorticity and a staggered flow consistent with SF order. To study
the degree of SF order we look at $\phi ({\v R} )$ of each plaquette. It is striking that for 
all three cases,
$\phi (\v R)$ is almost perfectly staggered for all the plaquettes shown. 
It is also clear that as one moves away from the center the magnitude of $\phi (\v R)$
decays. To quantify the degree of SF order we define the SF order parameter as the staggered sum of $\phi (\v R)$
in the central $5\times 5$ plaquettes of the lattice: \ba \Phi_{oafm}=\sum_{\v R} (-1)^R \phi(\v R). \ea The
results for $\Phi_{oafm}$ is shown in table I.

Figures 2(d)-(f) illustrate the Gutzwiller projected results for $x=6\%, 12\%, 18\%$ respectively.
The projection enhances the maximal current in the first two cases.
It is clear that there is still substantial amount of staggered order near the center of the vortex.
The values of the SF order parameter for these figures as well as their maximal currents are shown in table I.

By studying the value of $\Phi_{oafm}$ as shown in Table I we conclude that the  $x= 12\%$ system shows the strongest
evidence for the SF order in the vortex core. We should also stress that at still lower doping the vortex core
becomes insulating, and the bond current should vanish.

\widetext
\begin{figure}
\centering \hskip -0.2cm \epsfxsize=17cm \epsfysize=12cm
\epsfbox{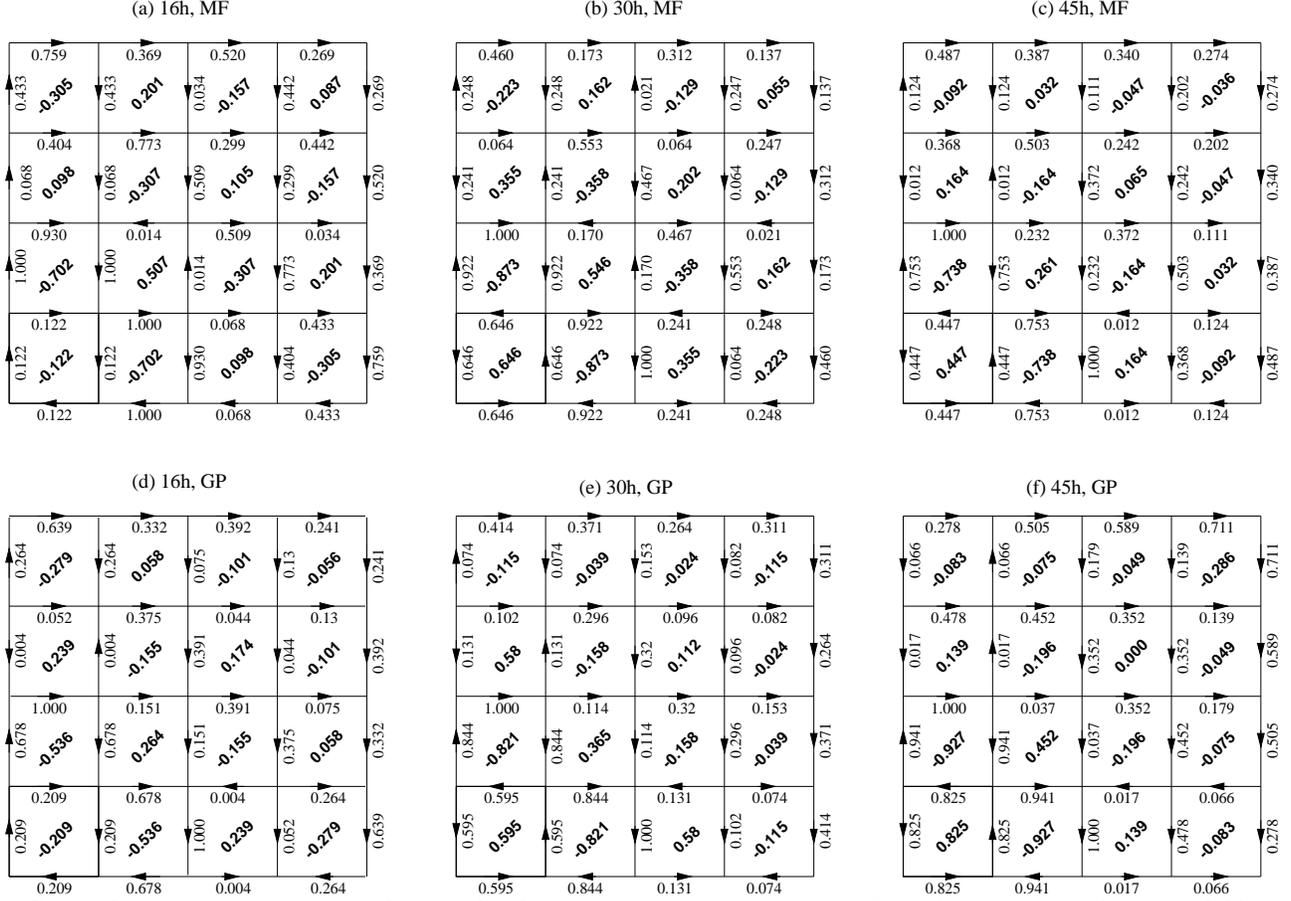} \caption{The current pattern near the
core for doping concentration $x=6\%, 12\%, 18\%$. (a)-(c) are the mean-field results and (d)-(f) are the results
after the Gutzwiller projection. Only one quadrant of the square lattice is shown. The center of the vortex is in
the middle of the lower-left-corner plaquette. Bond current, together with the direction of the flow is indicated.
The numbers are normalized by the maximal current.  The numerals in each plaquette is the lattice curl of the bond
currents. The maximum bond current are $|J_{ij}|_{max}=0.012, 0.039, 0.040$ for (a)-(c), and $0.023, 0.059, 0.036$ for (d)-(f).}
\end{figure}
\narrowtext

By comparing Fig. 1 and Fig. 2 it is clear that the emergence of staggered currents is a strong correlation
effect. An experimental detection of the SF core would strongly indicate that our understanding of the local
correlation in the cuprates based on the t-J model is qualitatively  correct.

How to detect the SF order in the vortex core is an important question. Lee and Wen suggested using the NMR as a
probe of the local magnetic field generated by the orbital currents. From an STM experimental point of view it is
important to ask whether the signature of the SF core can be decoded from the tunneling spectroscopy. For example
one might think that the doubling of unit cell associated with the SF order can give rise to real-space patterns
of the tunneling spectra suggestive of the presence of two sublattices.  Motivated by such a question we computed
the real space map of the tunneling density of state at a fixed voltage. The result shows a strong sublattice
dependence for a bias close to $\pm 0.1J$. This is shown in Figs. 3(a),(b) for $x=12\%$. In order to determine
whether such a sublattice dependence is indeed a consequence of the SF order we compute the same map for the BCS
vortex shown in Fig. 1. As is clear in Figs. 3(c),(d), the sublattice dependence is also present in the BCS
vortex. We are then forced to conclude that the sublattice dependence is not correlated with the SF order. Instead
we believe it is due to the nature of the quasiparticle wavefunctions near the gap nodes. We have also studied the
tunneling spectra as a function of bias voltage in a range of doping where the SF core is found. Despite the common presence of SF order, the detailed shape of the spectra depends sensitively on the doping concentration.
We therefore conclude that it is difficult to draw an unequivocal prediction for the  STM spectroscopy based on SF order.

In summary, we have studied in detail the current distribution near the vortex core both in the BCS and
strongly-correlated systems. We find in the latter case an evidence of staggered flux order near the optimum
doping. We take this as an indication that, 
next to the $d$-wave superconducting state, the staggered flux state is a
close contender for the ground state in the studied doping range. The strength of SF order weakens when the system
is doped away from optimal one. Combining the result of Ref.\cite{hl} and this work we have identified three
distinctive signatures of strong correlation in the core of a superconducting vortex. 1) There are two ways to suppress the superfluid density inside the vortex core, 
one by de-pairing and the other by locally depleting the
holes. They respectively give rise to the metallic and insulating vortex cores. 2) At energies below the charge gap, the single-particle spectral weight is not conserved. 3) The suppression of the paring in the metallic core results in the appearance of staggered flux order.

\begin{figure}
\centering \epsfxsize=8cm \epsfysize=8cm
\epsfbox{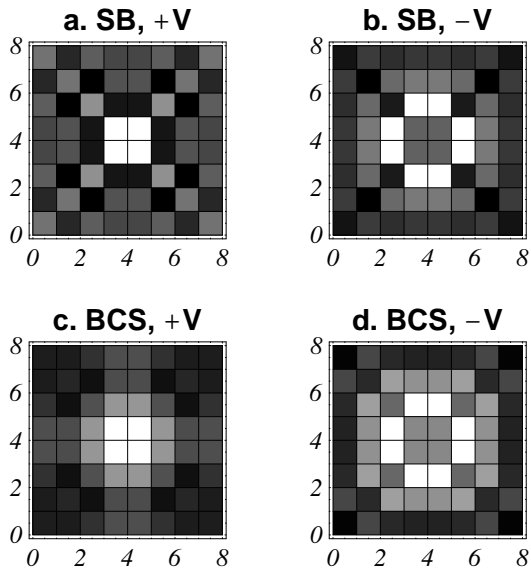}
\caption{A conductance profile at a fixed bias near  $\pm V$, $V\approx 0.1J$,
from slave-boson (SB)  mean-field theory ((a) and (b)) and
BCS-like (BCS)  mean-field theory ((c) and (d)) of the
vortex core.  The interior $8\times8$ region is shown. The average hole density in both cases is  $12\%$.}
\end{figure}
\noindent

{\bf Acknowledgment}\rm$~$ We acknowledge G. Baskaran, Seamus Davis, and Patrick Lee and for enlightening discussion, and Young-Gui Yoon for help with the figures.
DHL is supported in part by NSF grant DMR 99-71503. QHW is supported by the National Natural Science Foundation of China and the Ministry of Science and Technology of China (NKBSF-G19990646), and in part by the Berkeley Scholars Program. DHL and HJH are grateful to the ITP in Santa Barbara for generous hospitality and financial support by NSF grant No. PHY99-07949.

\begin{table}
\caption{Maximal Currents $|J_{ij}|$ and the staggered orbital magnetization $\Phi_{oafm}$ for various doping (See
text for exact definition). The superscripts MF/GP indicate mean-field and Gutzwiller projection results,
respectively.} \label{energy}
\begin{tabular}{cllll}
$x$  & $|J_{ij}|^{\mbox{MF}}$ & $|J_{ij}|^{\mbox{GP}}$ & $\Phi_{oafm}^{\mbox{MF}}$  & $\Phi_{oafm}^{\mbox{GP}}$ \\
\tableline
    6\% &  0.012 & 0.023 &  0.096    &   0.135  \\
  12\% &  0.039 & 0.059 &  0.445    &   0.553  \\
  18\% &  0.040 & 0.036 &  0.267    &   0.305  \\
\end{tabular}
 \end{table}

\widetext

\begin{references}
\bibitem{stripe} V. J.  Emery, S. A. Kivelson,  and J. M. Tranquada,
Proc. Natl. Acad. Sci. {\bf 96}, 8814 (1999).
\bibitem{ivanov} D. A.  Ivanov, P. A.  Lee, and  Xiao-Gang Wen,
Phys. Rev. Lett.  {\bf 84} 3958 (2000).
\bibitem{lt} J. L. Tallon and J. W. Loram, cond-mat/0005063.
\bibitem{clmn} S. Chakravarty, R. B. Laughlin, D. K. Morr and C. Nayak,
cond-mat/0005443.
\bibitem{leung}  P. W. Leung, cond-mat/0007068.
\bibitem{su2} Patrick  A. Lee  {\it et al.}  Phys. Rev. B {\bf 57}, 6003 (1998).
\bibitem{leewen} Patrick  A. Lee and Xiao-Gang Wen, cond-mat/0008419.
\bibitem{hl}  Jung Hoon Han and D.-H. Lee, Phys. Rev. Lett. {\bf 85}, 1100
(2000).
\bibitem{hwl} Jung Hoon Han, Qiang-Hua Wang, and D.-H. Lee, cond-mat/0006046.
\bibitem{sachdev} M. Vojta and S. Sachdev, Phys. Rev. Lett. {\bf 83}, 3916  (1999).
\bibitem{kotliar} G. Kotliar and J. Liu, Phys. Rev. B {\bf 38}, 5142 (1988).
\bibitem{affmst} I. Affleck and J. B. Marston, Phys. Rev. B {\bf 37},
3774 (1988).
\bibitem{boxphase} T. Dombre and G. Kotliar, Phys. Rev. B {\bf 39}, 855 (1989);
N. Read and S. Sachdev, Nucl. Phys. {\bf B316}, 609 (1989).
\bibitem{note} When antiferromagnetism is allowed the insulating vortex core is likely to exhibit local antiferromagnetic order.
\bibitem{note2} The BCS mean-field theory is obtained by
decoupling the antiferromagnetic interaction in Eq. \ref{h} in the $d$-wave pairing channel
and ignoring the occupation constraint.
\bibitem{yokoyama}  H. Yokoyama, and H. Shiba, J. Phys. Soc. Jpn {\bf 56},
3570 (1987).
\end{references}
\end{document}